\documentclass[a4paper,english,eqsecnum,a4paper,twocolumn,showpacs,preprintnumbers,amsmath,amssymb,prb]{revtex4}
\usepackage[latin1]{inputenc}
\usepackage{float}
\usepackage{graphicx}
\usepackage{amssymb}

\makeatletter

\usepackage{dcolumn}

\usepackage{bm}

\usepackage{babel}
\makeatother
\begin{document}

\title{Magnetic excitation 
in resonant inelastic x-ray scattering of Sr$_2$IrO$_4$: 
A localized spin picture}

\author{Jun-ichi Igarashi$^{1}$ and and Tatsuya Nagao$^{2}$}

\affiliation{
 $^{1}$Faculty of Science, Ibaraki University, Mito, Ibaraki 310-8512,
Japan\\
$^{2}$Faculty of Engineering, Gunma University, Kiryu, Gunma 376-8515,
Japan
}

\date{\today}

\begin{abstract}

We study the magnetic excitations in 5d transition-metal oxide
Sr$_2$IrO$_4$ on the basis of the Heisenberg model with small anisotropic
terms on a square lattice.
We calculate the correlation functions by using  the Green's functions
in the spin-wave approximation.
The spin waves are split into two modes with slightly different energies 
due to the anisotropic terms. It is shown that the spin correlation
functions of the $y$ and $z$ components are composed of a single peak 
corresponding to each mode.
We analyze the process of resonant inelastic x-ray scattering (RIXS)
without relying on the fast collision approximation to obtain the local
scattering operator. The RIXS intensity is derived as a sum of 
the correlation functions of the $y$ and $z$ spin components.
We demonstrate that the RIXS intensity as a function of energy 
shows two-peak structure brought about by the
two modes, which could be observed in the RIXS experiment.

\end{abstract}


\maketitle

\section{\label{sect.1}Introduction}

The $5d$ transition-metal compounds have recently drawn much attention
because of the interplay between the spin-orbit interaction (SOI) and 
the electron correlation.
Among them, Sr$_2$IrO$_4$ is one of the most fascinating systems 
due to the structural and electronic similarities to the La$_2$CuO$_4$,
parent compound of the high-$T_{\textrm{C}}$ superconductors.
This magnetic insulator, which exhibits a canted antiferromagnetic phase below $\sim$ 230 K, is proposed to be a system with an effective total angular momentum $j_{\textrm{eff}}=1/2$.
\cite{Crawford1994,Cao1998,Moon2006,Clancy2012,Ye2013,Dhital2013}
Many-body theoretical methods have been applied to the system
to describe the electronic structure.\cite{Kim2008,Arita2012,Watanabe2010}

In the strong coupling scheme, 
the localized electron picture may be useful 
to describe the low-lying excitations in such spin-orbit induced 
antiferromagnetic insulator.
This picture starts from the description of electronic states of
each Ir atom, where five $5d$ electrons in Ir$^{4+}$ ion 
are occupied in the $t_{2g}$  orbitals,
since the energy of the $e_g$ orbitals is about 2 eV higher than that of 
the $t_{2g}$  orbitals due to the strong crystal field.\cite{Kim2008} 
This situation may be regarded as one hole is sitting on the t$_{2g}$ orbitals. 
Under the strong SOI, the lowest-energy states of a hole are Kramers' 
doublet with $j_{\rm eff}=1/2$.\cite{Kim2008,Kim2009}

The degeneracy is lifted by the inter-site interaction.
Introducing the isospin operators acting on the doublet,
the effective spin Hamiltonian describing the low-lying excitations is 
derived by the second-order perturbation with respect to the electron-transfer 
terms, as has usually been carried out in the superexchange theory. 
\cite{Anderson1959}
A Heisenberg Hamiltonian is obtained with the isotropic 
antiferromagnetic coupling consistent with the above 
findings,\cite{Jackeli2009,Jin2009,Kim2012}
as well as small anisotropic terms, which arise when Hund's coupling
is taken into account on the two-hole states in the intermediate state of the 
second-order perturbation.\cite{Jackeli2009,Kim2012}
Since the anisotropic terms favor the staggered moment lying in the $ab$ plane,
the staggered moment is assumed to direct along the $x$ axis in the local
coordinate frames.
This leads to a zig-zag alignment of staggered moment along
the crystal $a$ axis, because the base states are defined in the
local coordinate frames rotated with respect to the $c$ axis 
about $\theta = \pm 11^{\circ}$ in accord with the rotation of 
the IrO$_6$ octahedra.\cite{Jackeli2009,Wang2011}  
On this situation, we have calculated the excitation
spectra within the linear spin-wave approximation\cite{Holstein1940} 
in our previous paper.\cite{Igarashi2013} 
Having introduced the Green's functions including 
the so-called anomalous type,\cite{Bulut1989} we have solved the coupled 
equations of motion for the Green's functions. We have found that magnon modes 
in the isotropic Heisenberg model are split into two modes with slightly 
different energy in the entire Brillouin zone, due to the anisotropic terms. 
This may be considered as a hallmark of the interplay between the SOI
and Hund's coupling.

Usually, inelastic neutron scattering (INS) works effective to
probe such magnetic excitations. However, it is not the case for 
this system, since the Ir atom is a strong absorber of neutron. 
On the other hand,
resonant inelastic x-ray scattering (RIXS) has recently emerged as
a useful probe for detecting the magnetic excitations. It has detected 
the single-magnon excitations as well as the two-magnon excitations 
in undoped cuprates, where the spectral peak behaves like the dispersion 
relation of spin wave in the Heisenberg model as a function of momentum 
transfer.\cite{Braicovich2009,Braicovich2010,Guarise2010}
For Sr$_2$IrO$_4$, the RIXS experiments have also been carried 
out around the Ir L$_3$ edge.\cite{Ishii2011,J.Kim2012}
A low-energy peak arising from one-magnon excitations 
has been observed similar to undoped cuprates,\cite{J.Kim2012} 
but no indication of the mode splitting has been seen. 
The theoretical analysis in the spin-wave approximation has been carried out,
having described well the spectra, but without considering the anisotropic
terms.\cite{Ament2011} At present, it is not clear how the split modes due 
to the anisotropic terms could be observed in the RIXS spectra.
The purpose of this paper is to clarify the origin of two modes and
how they are detected.

To this end, we introduce a pair of combination of spin operators
${\bf S}_a \pm {\bf S}_b$ where ${\bf S}_a$ and ${\bf S}_b$ are
the spin operators at A and B sites, respectively.
We call ${\bf S}_a + {\bf S}_b$ and ${\bf S}_a - {\bf S}_b$
as bonding and antibonding combinations, respectively. We consider 
the correlation functions of them, which 
are connected to the Green's functions mentioned above.
Since the staggered moment aligns along
the $x$ axis, the correlation function of the $x$ spin component, 
which consists of two-magnon excitations, could be neglected
as a higher order correction of $1/S$ expansion.
We find that the bonding-combination functions of the $y$ and $z$ spin 
components consist of a single $\delta$-function peak corresponding to 
each mode. The modes corresponding to the $y$ and $z$ spin
components are interchanged in the antibonding-combination functions.

These correlation functions are combined in evaluating the RIXS spectra. 
Analyzing the second-order RIXS process similar to the case 
for undoped cuprates,\cite{Igarashi2012}
we obtain the expression of the local scattering operator
described in terms of the spin operators at the core-hole site. 
The operator consists of a term consistent with the fast collision
approximation (FCA)\cite{Ament2007,Ament2009,Haverkort2010} and an extra
term not given by FCA. However, since the lifetime 
broadening width is larger than the magnon energies at the Ir L edge, 
the latter term is considered quite 
small and could be neglected in the present system.  
Using the local scattering operator derived, we can express
the RIXS spectra as a sum of the correlation functions of $y$ and $z$ 
spin components. Since the $\delta$-function peak energy is different between the functions with the spin components, 
the RIXS spectra are made up of two peaks. 
We also find that the correlation function of the
\emph{antibonding} spin combination for the momentum transfer
${\bf q}$ \emph{inside} the magnetic Brillouin zone (MBZ)
leads to the divergence of the intensity at ${\bf q}=(0,0)$.
It attributes to the zig-zag arrangement of the staggered 
moment, which is a consequence of the rotation of IrO$_6$ octahedra.
Evaluating the spectra in the model with the reasonable parameter values,
we demonstrate that the mode splitting could be distinguished.

This paper is organized as follows.
In Sec. \ref{sect.2}, we introduce the spin Hamiltonian with anisotropic terms 
in the square lattice.
The excitation spectra are calculated in the spin-wave approximation
with the help of the Green's functions. The correlation functions are evaluated
for the bonding and antibonding spin combinations.
In Sec. \ref{sect.3}, the RIXS process is analyzed at a single site
without relying on the FCA. In Sec. \ref{sect.4}, the RIXS spectra are
calculated for Sr$_2$IrO$_4$. Section \ref{sect.5} is devoted to the 
concluding remarks.
In Appendix, the symmetry relations among the Green's functions are
summarized.

\section{\label{sect.2}Magnetic excitations for 
S\lowercase{r}$_2$I\lowercase{r}O$_4$}

\subsection{Spin Hamiltonian}
The crystal structure of Sr$_2$IrO$_4$ belongs to the K$_2$NiF$_4$ type.
\cite{Crawford1994}
The IrO$_2$-layer forms two-dimensional plane 
similar to the CuO$_2$-layer in La$_2$CuO$_4$.
The crystal field energy of the $e_g$ orbitals is about 2 eV higher than 
that of the $t_{2g}$ orbitals.
This yields five electrons to be occupied on $t_{2g}$ 
orbitals in each Ir atoms.
This state could be considered as occupying one \emph{hole}.
The matrices of the orbital angular momentum operators with $L=2$ represented 
by the $t_{2g}$ states are the minus of those with $L=1$ represented by 
$|p_x\rangle$, $|p_y\rangle$, and $|p_z\rangle$, 
if the bases are identified by
$|yz\rangle$, $|zx\rangle$, and $|xy\rangle$, 
respectively.\cite{Kanamori1957}
Therefore, the six-fold degenerate states are split 
into the states with the effective angular momentum $j_{\rm eff}=1/2$ and
$3/2$ under SOI.
The lowest-energy states are the doublet with $j_{\rm eff}=1/2$, given by
\begin{eqnarray}
 \left|\uparrow \right\rangle &=& \frac{1}{\sqrt{3}}\left[
  |yz\downarrow \rangle + i|zx\downarrow\rangle +|xy\uparrow \rangle\right],\\
 \left|\downarrow \right\rangle &=& \frac{1}{\sqrt{3}}\left[
   |yz\uparrow\rangle - i|zx\uparrow\rangle -|xy\downarrow \rangle \right].
\end{eqnarray}
where the base states are defined in the local coordinate frames rotated 
in accordance with the rotation of the IrO$_6$ octahedra.
\cite{Jackeli2009,Wang2011} 

The exchange interaction with neighboring doublets is evaluated from the 
perturbation with respect to the electron transfer in the strong coupling 
theory.\cite{Jackeli2009,Kim2012}
By introducing the spin operators ${\bf S}$ acting on the doublet, 
the effective Hamiltonian may be 
expressed as   
\begin{equation}
 H= H^{(0)} + H^{(1)},
\end{equation}
with
\begin{eqnarray}
 H^{(0)}&=& J_{\rm ex}\sum_{\langle i,j\rangle}{\bf S}_i\cdot{\bf S}_j 
 +J'_{\rm ex}\sum_{\langle i',j'\rangle}{\bf S}_{i'}\cdot{\bf S}_{j'} 
\nonumber \\
 &+&
J''_{\rm ex}\sum_{\langle i'',j''\rangle}{\bf S}_{i''}\cdot{\bf S}_{j''}
+ \cdots, \\
 H^{(1)}&=&J'_z \sum_{\langle i,j\rangle} S_i^z S_j^z
         +J'_{xy}\sum_{\langle i,j\rangle} \textrm{sgn}(i,j)
          \left(S_i^x S_j^x - S_i^y S_j^y\right). 
\nonumber \\
\label{eq.H.1}
\end{eqnarray}
The $H^{(0)}$ describes the isotropic exchange energy where
the exchange couplings between the first, second, and third nearest-neighbors are denoted as 
$J_{\rm ex}$, $J'_{\rm ex}$, and $J''_{\rm ex}$, respectively. 
The summations $\langle i,j \rangle$, $\langle i',j' \rangle$, and
$\langle i'',j'' \rangle$ run over the first, second, and third nearest-neighbor pairs, respectively.
It is known that the experimental dispersion curve can be 
reproduced well by setting $J_{\textrm{ex}}=60$ meV, 
$J_{\textrm{ex}}'=-J_{\textbf{ex}}/3$ and 
$J_{\textrm{ex}}''=J_{\textbf{ex}}/4$ in the phenomenological model.
\cite{J.Kim2012}
The $H^{(1)}$ describes the anisotropic exchange energy, which arises
from the interplay between the SOI and Hund's coupling, 
where $\textrm{sgn}(i,j)$ gives $+1 (-1)$ when the bond 
between the sites $i$ and $j$ is along the $x$ ($y$) axis. 
It is known that  $J'_{z}$ is negative and 
its absolute value is nearly the same as that of $J'_{xy}$.
\cite{Jackeli2009,Kim2012,Igarashi2013} 
It may be sufficient to restrict the anisotropic interaction within 
the nearest neighbors, since it is one order of magnitude smaller than
the isotropic term. 

\subsection{The ground state} 
In the absence of the anisotropic term $H^{(1)}$, the conventional antiferromagnetic spin configuration is expected, in which
the direction of the staggered moment is not determined. 
The first term of $H^{(1)}$ makes the direction favor 
the $xy$ plane when $J'_z < 0$.
This antiferromagnetic order breaks the rotational invariance of the isospin
space in the $ab$ plane.
We assume the staggered moment pointing to the $x$ axis.\cite{Cao1998}
It should be noted here that the antiferromagnetic order in the local coordinate
frames indicates a zig-zag alignment of the staggered moment, leading to 
the presence of the weak ferromagnetic moment in the coordinate frame
fixed to the crystal axes.

\subsection{Excited states}
A spin-wave theory has been developed to describe excited states in Ref.
\onlinecite{Igarashi2013}.
Relabeling the $x$, $y$, and $z$ axes as $z'$, $x'$, and $y'$ 
axes, respectively, we express the spin operators by boson operators 
within the lowest order of $1/S$-expansion:\cite{Holstein1940}
\begin{eqnarray}
 S_i^{z'} &=& S - a_i^\dagger a_i ,  \quad 
 S_i^{x'}+iS_i^{y'} = \sqrt{2S}a_i , \label{eq.boson1}\\
 S_j^{z'} &=& -S + b_j^\dagger b_j , \quad
 S_j^{x'}+iS_j^{y'} = \sqrt{2S}b_j^\dagger ,\label{eq.boson2}
\end{eqnarray}
where $a_i$ and $b_j$ are boson annihilation operators,
and $i$ ($j$) refers to sites on the A (B) sublattice. 
Then, the Fourier transforms of spin operators are defined 
in the MBZ as 
\begin{eqnarray}
 {\bf S}_a({\bf k}) &=& \sqrt{\frac{2}{N}}\sum_{i}{\bf S}_{i}
                             \exp(-i{\bf k}\cdot{\bf r}_i) , \\
 {\bf S}_b({\bf k}) &=& \sqrt{\frac{2}{N}}\sum_{j}{\bf S}_{j}
                             \exp(-i{\bf k}\cdot{\bf r}_j) ,
\label{eq.Fouriers2}
\end{eqnarray}
where $N$ is the number of sites, and $i$ ($j$) runs over A (B) sublattice.
Defining similarly the Fourier transform of boson operators $a({\bf k})$
and $b({\bf k})$, we express the Hamiltonian as  
\begin{eqnarray}
&& H^{(0)} = J_{\rm ex}Sz\sum_{\bf k} \left\{
a^{\dagger}({\bf k})a({\bf k}) 
  + b^{\dagger}({\bf k})b({\bf k}) \right. \nonumber \\
& & \left. + 
 \gamma({\bf k})[a^{\dagger}({\bf k})b^{\dagger}({\bf -k})
                   +a({\bf k})b({\bf -k})] \right\} \nonumber \\
             &-& J'_{\rm ex}Sz\sum_{\bf k} [1-\gamma'({\bf k})]
         [a^{\dagger}({\bf k})a({\bf k}) + b^{\dagger}({\bf k})b({\bf k})] \nonumber\\
        &-& J''_{\rm ex}Sz\sum_{\bf k} [1-\gamma''({\bf k})]
         [a^{\dagger}({\bf k})a({\bf k}) + b^{\dagger}({\bf k})b({\bf k})] , \\ 
&& H^{(1)} = J'_{z}(2S)\sum_{\bf k} \gamma({\bf k})
    [a({\bf k})-a^{\dagger}({\bf -k})][b({\bf -k}) - b^{\dagger}({\bf k})]\nonumber \\
           &-& J'_{xy}(2S)\sum_{\bf k} \eta({\bf k})
         [a({\bf k})+a^{\dagger}({\bf -k})][b({\bf -k})+b^{\dagger}({\bf k})], 
\end{eqnarray}
where 
\begin{eqnarray}
 \gamma({\bf k}) &=& \frac{1}{2}(\cos k_x + \cos k_y), \\
 \gamma'(\bf{k}) &=& \cos k_x \cos k_y, \\
\gamma''(\bf{k}) &=& \frac{1}{2} [ \cos(2k_x) + \cos(2 k_y)], \\
 \eta({\bf k}) &=& \frac{1}{2}(\cos k_x - \cos k_y).
\end{eqnarray}
Here $z$ is the number of nearest neighbors, i.e., $z=4$.

To find out the excitation modes, we introduce the Green's functions,
\begin{eqnarray}
 G_{aa}({\bf k},t) &=& -i\langle T[a({\bf k},t)a^{\dagger}({\bf k},0)]\rangle, \\
 F_{ba}({\bf k},t) &=& -i\langle T[b^{\dagger}({\bf -k},t)a^{\dagger}({\bf k},0)]\rangle, \\
 G_{ba}({\bf k},t) &=& -i\langle T[b({\bf k},t)a^{\dagger}({\bf k},0)]\rangle, \\
 F_{aa}({\bf k},t) &=& -i\langle T[a^{\dagger}({\bf -k},t)a^{\dagger}({\bf k},0)]\rangle,
\end{eqnarray}
where $T$ is a time-ordering operator, and $\langle X \rangle$ denotes the ground-state
average of operator $X$. The $F_{ba}({\bf k},t)$ and $F_{aa}({\bf k},t)$ belong to the 
so called anomalous type. Their Fourier transforms are defined as 
$G_{aa}({\bf k},\omega) = \int G_{aa}({\bf k},t){\rm e}^{i\omega t}{\rm d}t$
and so on. Then, we get a set of equation 
of motion for these functions. It is given by 
\begin{widetext}
\begin{equation}
 \left( \begin{array}{cccc}
  \omega -1+\xi(\textbf{k}) & -A({\bf k}) & B({\bf k}) & 0 \\
  -A({\bf k})  & -(\omega+1-\xi(\textbf{k})) & 0 & B({\bf k}) \\
  B({\bf k}) & 0 & \omega-1+\xi(\textbf{k}) & -A({\bf k}) \\
  0 & B({\bf k}) & -A({\bf k}) & -(\omega+1-\xi(\textbf{k})) 
  \end{array} \right)
  \left( \begin{array}{c}
         G_{aa}({\bf k},\omega) \\
         F_{ba}({\bf k},\omega) \\
         G_{ba}({\bf k},\omega) \\
         F_{aa}({\bf k},\omega)
        \end{array} \right)
  = \left( \begin{array}{c}
            1 \\
            0 \\
            0 \\
            0 
           \end{array} \right) , \label{eq.matrix}
\end{equation}
\end{widetext}
where 
\begin{eqnarray}
\xi(\textbf{k})&=& \frac{J_{\textrm{ex}}'}{J_{\textrm{ex}}}
( 1-\gamma'(\textbf{k}))
+ \frac{J_{\textrm{ex}}''}{J_{\textrm{ex}}}( 1-\gamma''(\textbf{k})), \\
   A({\bf k}) &=& (1+g_z)\gamma({\bf k}) -g_{xy}\eta({\bf k}) , \\
   B({\bf k}) &=& g_z\gamma({\bf k}) + g_{xy}\eta({\bf k}) , \\
   g_z &=& \frac{J'_{z}}{2J_{\rm ex}}, \quad g_{xy}=\frac{J'_{xy}}{2J_{\rm ex}}. 
\end{eqnarray}
Here the energy is measured in units of $J_{\rm ex}Sz$.
Hence we finally obtain,
\begin{equation}
\left( \begin{array}{c}
 G_{aa}({\bf k},\omega) \\
 F_{ba}({\bf k},\omega) \\
 G_{ba}({\bf k},\omega) \\
 F_{aa}({\bf k},\omega) \\
\end{array} \right)=\frac{1}{D({\bf k},\omega)}
\left( \begin{array}{c}
g_{aa}({\bf k},\omega) \\
f_{ba}({\bf k},\omega) \\
g_{ba}({\bf k},\omega) \\
f_{aa}({\bf k},\omega) \\
\end{array} \right), 
\label{eq.sol}
\end{equation}
where
\begin{eqnarray}
 g_{aa}({\bf k},\omega) &=& [\omega-1+\xi({\bf k})][\omega+1-\xi({\bf k})]^2 
\nonumber \\
 &-& B({\bf k})^2[\omega-1+\xi({\bf k})]
 +A({\bf k})^2[\omega+1-\xi({\bf k})] , \nonumber \\
 \\
 f_{ba}({\bf k},\omega) &=& -A({\bf k})\{\omega^2-[1-\xi({\bf k})]^2
-B({\bf k})^2 +A({\bf k})^2\}, \nonumber \\
\\
 g_{ba}({\bf k},\omega) &=& B({\bf k})
      \{B({\bf k})^2-[\omega+1-\xi({\bf k})]^2-A({\bf k})^2\}, \nonumber \\
\\
 f_{aa}({\bf k},\omega) &=& 2A({\bf k})B({\bf k})[1-\xi({\bf k})].
\label{eq.faa}
\end{eqnarray}
The denominator of Eq. (\ref{eq.sol}) is given by
\begin{equation}
D(\textbf{k},\omega) =[ \omega^2 - E_-^2(\textbf{k})]
[\omega^2 - E_+^2(\textbf{k})],
\end{equation}
with
\begin{equation}
 E_{\pm}(\textbf{k}) = 
\sqrt{ [1 - \xi(\textbf{k})
\pm |B(\textbf{k})|]^2 - A^2(\textbf{k}) }. \label{eq.disp.2}
\end{equation}
This indicates that poles exist at $\omega=E_{\pm}({\bf k})$ 
in the domain of $\omega> 0$. When ${\bf k}\to 0$, we have $\xi({\bf k})\to 0$,
$A({\bf k})\to 1+g_z$, and $B({\bf k})\to g_z$, which leads to a
Goldstone mode $E_{-}(0)=0$ as well as a gap mode $E_{+}(0)=\sqrt{-2g_z}$. 
The splitting of two modes is a direct reflection of the anisotropy
shown in the original Hamiltonian (\ref{eq.H.1}).
Note that since $B(\textbf{k})$ is not invariant under the 
exchange of $k_x$ and $k_y$, the dispersion shows a slight anisotropy
though the difference is negligible due to the smallness of
$J_z '$ and $J_{xy}'$ in the following numerical evaluations.

Finally, evaluating the residues at the poles, we could express the Green's 
function, for example, $G_{aa}({\bf k},\omega)$ as
\begin{equation}
 G_{aa}({\bf k},\omega) = \sum_{\mu=\pm}\left\{
 \frac{A_{\mu}}{\omega-E_{\mu}({\bf k})+i\delta}
-\frac{B_{\mu}}{\omega+E_{\mu}({\bf k})-i\delta}\right\},
\label{eq.greensfunction}
\end{equation}
where $\delta$ is an infinitesimal positive constant.
The Green's functions are utilized when we evaluate the spin correlation
functions in the next subsection.

\subsection{Spin correlation function}
Since two spins exist in the unit cell, it is useful to define
a pair of combination of spin operators 
\begin{equation}
 {\bf Q}_{\pm}({\bf k}) \equiv \frac{1}{\sqrt{2}}
[{\bf S}_a({\bf k}) \pm {\bf S}_b({\bf k})],
\end{equation}
where $\textbf{Q}_+(\textbf{k})$ and $\textbf{Q}_-(\textbf{k})$ 
are called as bonding and antibonding
combinations, respectively.
The antibonding combination corresponds to the wave vector 
${\bf k}'={\bf k}+{\bf G}$ outside 
the first MBZ in the extended zone scheme, since ${\bf S}_b({\bf k}')$ 
acquires a minus sign when it is reduced back to the first MBZ by 
a reciprocal lattice vector ${\bf G}$. 

The INS and RIXS spectra may be connected to
the correlation functions of these operators,
\begin{equation}
 R_{\ell}^{\mu\mu}({\bf k},\omega) = \int 
 \langle Q_{\ell}^{\mu}({\bf k},t)Q_{\ell}^{\mu}(-{\bf k},0) 
 \rangle {\rm e}^{i\omega t}{\rm d}t, 
\label{eq.R}
\end{equation}
with $\mu=x, y$, and $z$. Since the direction of the staggered moment is
along the $x$ axis, the $R_{\ell}^{xx}({\bf k},\omega)$, composed of 
two-magnon excitations, is regarded as the higher order of the $1/S$ 
expansion, and will be neglected.  
The $R_{\ell}^{yy}({\bf k},\omega)$ and $R_{\ell}^{zz}({\bf k},\omega)$,
composed of one-magnon excitations, are different with each other
because of the anisotropic terms of $J'_{z}$ and $J'_{xy}$.
To evaluate these functions, we decompose the right hand side 
of Eq.~(\ref{eq.R}) into the correlation functions of Holstein-Primakoff 
bosons such as 
$\int\langle b({\bf k},t)a^{\dagger}({\bf k},0)\rangle{\rm e}^{i\omega t}
{\rm d}t$, and connect them to the imaginary part of the Green's functions
such as $-2{\rm Im}\,G_{ba}({\bf k},\omega)$ for $\omega>0$. 
All the Green's functions required are obtained from Eq.~(\ref{eq.sol}) 
with the help of the symmetry relations given in Appendix.

From the form of Eq. (\ref{eq.greensfunction}), we see that
each correlation function has a single $\delta$-function peak structure.
For instance, for $\textbf{k}$ from $(0,0)$ to $(\pi,0)$, 
we find that $R_{+}^{zz}({\bf k},\omega)$ and 
$R_{-}^{zz}({\bf k},\omega)$ are 
composed of the $\delta$-function peaks at $E_{-}({\bf k})$
and $E_{+}({\bf k})$, respectively. 
On the other hand, $R_{+}^{yy}({\bf k},\omega)$ and
$R_{-}^{yy}({\bf k},\omega)$ are composed 
of the peaks at $E_{+}({\bf k})$ and $E_{-}({\bf k})$, respectively. 
When we turn our attention to the diagonal direction of $\textbf{k}$,
it is convenient to modify the definition of the correlation functions
in the extended zone scheme. 
Since the antibonding combination corresponds 
to the wave number belonging to the outside of the first MBZ, 
we define the correlation functions by
$R^{\mu\mu}({\bf k},\omega)\equiv R_{+}^{\mu\mu}({\bf k},\omega)$ 
for ${\bf k}$ inside the first MBZ, and 
$R^{\mu\mu}({\bf k},\omega)\equiv R_{-}^{\mu\mu}([{\bf k}],\omega)$ 
for ${\bf k}$ outside the 1st MBZ where 
$[\textbf{k}]$ is the wave vector reduced back to the first MBZ by 
a reciprocal lattice vector as ${\bf G}$$=\textbf{k} - [\textbf{k}]$.

\begin{figure}
\includegraphics[width=8.0cm]{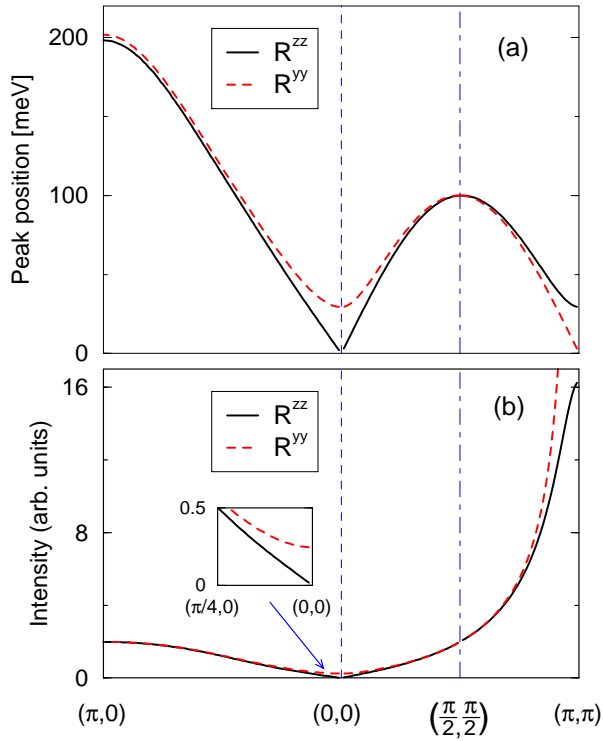}%
\caption{\label{fig.corr} (Color online)
(a)The $\delta$-function peak positions for $R^{zz}({\bf k},\omega)$
(solid line) and $R^{yy}({\bf k},\omega)$ (broken line) for ${\bf k}$
along symmetry lines.
(b)Intensities of the peak for $R^{zz}({\bf k},\omega)$
and for $R^{yy}({\bf k},\omega)$. The gap mode at $(0,0)$ has a finite
intensity as shown in Inset.
}
\end{figure}

In the numerical calculation, we use the parameter values,
$J_{\rm ex}=60$, $J'_{\rm ex}=-20$, $J_{\rm ex}''=15$, $J'_{z}=-1.8$,
and $J'_{xy}=1.8$ in units of meV.
%%%%%%%%%%%%  revised %%%%%%%%%%%%
The parameter set used here is the same as the one adopted in
Ref. \onlinecite{Igarashi2013}, 
which is justified to give a better fitting of the
dispersion curve of the magnetic excitation obtained by the RIXS 
experiment.\cite{J.Kim2012} 
Notice that the magnitudes of the anisotropic exchange couplings 
$J_z '$ and $J_{xy}'$ turn out to be the same order as those evaluated
by other theories.\cite{Jackeli2009,Kim2012}
%%%%%%%%%%%%%%%%%%%%%%%%%%%%%%%%%%%%%%%%%%
Panel (a) in Fig. \ref{fig.corr} shows the peak positions 
of $R^{zz}({\bf k},\omega)$ and $R^{yy}({\bf k},\omega)$ 
as a function of ${\bf k}$ along symmetry lines.
The peak position of $R^{zz}({\bf k},\omega)$ has no gap at $(0,0)$ but 
has a gap at $(\pi,\pi)$, while the situation is opposite for
the peak of $R^{yy}({\bf k},\omega)$.
Panel (b) in Fig. \ref{fig.corr} shows the intensities of the peaks.
The intensity of $R^{zz}({\bf k},\omega)$ vanishes at  ${\bf k}=(0,0)$, 
and grows large but remains finite around ${\bf k}=(\pi,\pi)$.
The intensity of $R^{yy}({\bf k},\omega)$ remains finite but is 
quite small for ${\bf k}=(0,0)$, and diverges at ${\bf k}=(\pi,\pi)$.  
\\

\section{\label{sect.3}Scattering operator of RIXS at the $L_{2,3}$ edge}
\subsection{Second-order optical process}
The RIXS process is described by the electron-photon interaction
Hamiltonian $H_{\textrm{int}}$.
In the second-order optical process, the incident
photon with wave vector $\textbf{q}_i$, energy $\omega_i$, and
polarization $\alpha_i$ is absorbed by the material system, then
the scattered photon with wave vector $\textbf{q}_f$, 
energy $\omega_f$, and polarization $\alpha_f$ is emitted.
Then, the RIXS intensity $W(q_f\alpha_f;q_i\alpha_i)$ is written as,
\begin{eqnarray}
 W(q_f\alpha_f;q_i\alpha_i) 
 &=& 2\pi\sum_{f}\left|\sum_{n}
  \frac{\langle \Phi_f|H_{\rm int}|n\rangle \langle n|H_{\rm int}
 |\Phi_i\rangle}
{E_g+\omega_i-E_n} \right|^2 \nonumber\\
 &\times&\delta(E_g+\omega_i-E_{f}-\omega_f).
\label{eq.optical}
\end{eqnarray} 
where $q_i\equiv({\bf q}_i,\omega_i)$ and $q_f\equiv({\bf q}_f,\omega_f)$. 
The initial and final states are given by
$|\Phi_{i}\rangle=c_{q_i \alpha_i}^\dagger|g\rangle|0\rangle$
and $|\Phi_{f}\rangle=c_{q_f \alpha_f}^\dagger|f\rangle|0\rangle$,
respectively, 
where $|g\rangle$ and $|f\rangle$ represent the ground and excited states
of the matter with energies $E_{g}$ and $E_{f}$,
respectively.
The creation (annihilation) operator of the photon 
is denoted as $c_{\textbf{q} \alpha}^{\dagger}$ 
($c_{\textbf{q} \alpha}$), which acts on
the photon vacuum $|0\rangle$.
The intermediate state $|n\rangle$ represents 
the eigenstate of the matter with energy $E_n$ in the presence of 
core hole.

At the Ir $L_{2,3}$ edge, $H_{\rm int}$ represents 
the electric dipole ($E$1) transition
where a $2p$-core electron is excited to 
the $5d$ states. By restricting the transition
within the manifold of $j_{\rm eff}=1/2$, it may be expressed as
\begin{equation}
H_{\rm int}=w\sum_{\textbf{q}}
\frac{1}{\sqrt{2\omega_{\textbf{q}}}}
\sum_{i, m, \sigma}D^{\alpha}(jm,\sigma)
h_{i,\sigma}p_{i,jm}c_{\textbf{q}\alpha}
{\rm e}^{i\textbf{q}\cdot\textbf{r}_{i}} +{\rm H.c.},
\label{eq.tran1}
\end{equation}
where $w$ is a constant proportional to 
$\int_0^{\infty}r^3R_{5d}(r)R_{2p}(r){\rm d}r$,
with $R_{5d}(r)$ and $R_{2p}(r)$ being the radial wave-functions for 
the $5d$ and $2p$ states of Ir atom. 
The $p_{jm}$ ($p_{jm}^{\dagger}$) stands for the annihilation 
(creation) operator of the $2p$ 
core electron with the angular momentum $jm$,
which states are defined in the local crystal coordinate frame. 
The operator $h_{i,\sigma}$ ($h_{i,\sigma}^{\dagger}$) represents the annihilation 
(creation) of 5d hole at site $i$ 
with the Kramers' doublet specified by $\sigma$ 
($=\uparrow$ or $\downarrow$)
in hole picture, which quantization axis is rotated from the local crystal 
coordinate frame with Euler angles 
$\alpha$, $\beta$, and $\gamma$.\cite{Com1}
The coefficient $D^{\alpha}(jm,\sigma)$ describes the dependence on 
the $5d$ and core-hole states, which can be calculated in 
a similar manner as explained in Ref. \onlinecite{Igarashi2012}
for cuprates.

\subsection{Excitation and deexcitation of core hole at a single site}
We analyze the situation that the core electron is excited and deexcited
at the origin by following the procedure developed for undoped cuprates.
The intermediate state
just after the $E1$ transition takes place is given by  
\begin{equation}
 H_{\rm int}|g\rangle \propto\sum_{m}\left[ 
\sum_{\sigma=\uparrow,\downarrow}
    D^{\alpha_i}(jm,\sigma)|\psi_0^{\sigma}\rangle
  \right] |jm\rangle.
\end{equation}
Here we write $|g\rangle$ as
\begin{equation}
 |g\rangle = |\uparrow\rangle|\psi_0^{\uparrow}\rangle
           + |\downarrow\rangle|\psi_0^{\downarrow}\rangle,
\end{equation}
where $|\uparrow\rangle$ and $|\downarrow\rangle$ represent the normalized 
spin states at the origin, while $|\psi_0^{\uparrow}\rangle$ and 
$|\psi_0^{\downarrow}\rangle$ are constructed by the bases of the rest of 
spins, which are not normalized.
The core hole state is represented as $|jm\rangle$.
Note that the spin degrees of freedom of the $5d$ state 
is lost at the core-hole site,
which is reminiscent of the introduction of non-magnetic impurity
into spin system.
Employing the normalized eigenstate $|\phi_{\eta}\rangle$'s 
with eigenvalue $\epsilon'_{\eta}$ in the intermediate state, we have  
\begin{eqnarray}
 |F\rangle &\equiv& \sum_n H_{\rm int}|n\rangle\frac{1}{\omega_i+E_g-E_n}
   \langle n|H_{\rm int}|\Phi_i\rangle \nonumber \\
   &\propto& \sum_{m, \sigma, \sigma'} D^{\alpha_f}(jm,\sigma)^{*}
                             D^{\alpha_i}(jm,\sigma') 
   \nonumber \\
&\times& \sum_{\eta}|\sigma\rangle |\phi_{\eta}\rangle R(\epsilon'_{\eta})
    \langle\phi_{\eta}|\psi_{0}^{\sigma'}\rangle,
\label{eq.process1}
\end{eqnarray}
with
\begin{equation}
 R(\epsilon'_{\eta}) =
 \frac{1}{\omega_i+\epsilon_g -\epsilon_{\rm core}+i\Gamma - \epsilon'_{\eta}},
\end{equation}
where $\epsilon_{g}$ and $\epsilon_{\rm core}$ denote the ground 
state energy of the magnetic system and 
the energy required to create a core hole in the state 
$|jm\rangle$ and the $5d^{6}$-configuration, respectively.
The life-time broadening width of the core hole is denoted 
as $\Gamma$, which is around a few eV at the $L$ edge.\cite{Krause1979,Clancy2012}
The first factor in the right hand
side of Eq.~(\ref{eq.process1}) is rewritten as
\begin{eqnarray}
\sum_{m}D^{\alpha_f}(jm,\sigma)^{*}D^{\alpha_i}(jm,\sigma)
&\equiv& P_{\sigma}^{(0)}(j;\alpha_f,\alpha_i), \\
\sum_{m}D^{\alpha_f}(jm,\sigma)^{*}D^{\alpha_i}(jm,-\sigma) 
&\equiv& P_{\sigma}^{(1)}(j;\alpha_f,\alpha_i),
\end{eqnarray}
where $-\sigma$ denotes $\downarrow$ for $\sigma=\uparrow$
and vice versa. The $P_{\sigma}^{(0)}$ and $P_{\sigma}^{(1)}$ correspond to 
the spin-conserving and the spin-flip processes, respectively, whose 
values for $j=\frac{3}{2}$ are listed in Table \ref{table.1}
for $\alpha_i$ and $\alpha_f$ along the $x$, $y$, and $z$ axes.
Note that they retain finite values even for the $z$ polarization, 
which contrasts with the case of the undoped cuprates where 
the $z$ polarization has no finite contribution.\cite{Igarashi2012}
It can be confirmed that they vanish for $j=\frac{1}{2}$, consistent 
with the $L_{2}$ absorption experiment.\cite{Kim2009}

\begin{table*}
\caption{\label{table.1}
$P_{\sigma}^{(0)}(\frac{3}{2};\alpha_f,\alpha_i)$ and 
$P_{\sigma}^{(1)}(\frac{3}{2};\alpha_f,\alpha_i)$ 
where upper and lower signs correspond to $\sigma=\uparrow$ and $\downarrow$, 
respectively.
}
\begin{ruledtabular}
\begin{tabular}{cccc}
      & & $P_{\sigma}^{(0)}$ &  \\
\hline
   $\alpha_f \setminus \alpha_i$ & $x$  & $y$ & $z$ \\ 
\hline
 $x$ & $\frac{2}{15}$ & $\mp \frac{i}{15}\cos\beta$ 
& $\pm \frac{i}{15}\sin\alpha\sin\beta$ \\
 $y$ & $\pm\frac{i}{15}\cos\beta$ & $\frac{2}{15}$
& $\mp \frac{i}{15}\cos\alpha\sin\beta$ \\
 $z$ & $\mp \frac{i}{15}\sin\alpha\sin\beta$ 
& $\pm \frac{i}{15}\cos\alpha\sin\beta$ & $\frac{2}{15}$ \\
\hline
      & & $P_{\sigma}^{(1)}$ & \\
\hline
 $x$ & $0$ & $\frac{i}{15}\sin\beta\textrm{e}^{\pm i \gamma}$
  & $\pm\frac{1}{15}[\cos\alpha\pm i\sin\alpha\cos\beta] 
    \textrm{e}^{\pm i \gamma}$ \\
 $y$ & $-\frac{i}{15}\sin\beta\textrm{e}^{\pm i \gamma}$ & $0$ 
  & $-\frac{i}{15}[\cos\alpha\cos\beta \pm i\sin\alpha]
 \textrm{e}^{\pm i \gamma}$  \\
 $z$ & $\mp\frac{1}{15}[\cos\alpha \pm i\sin\alpha\cos\beta]
 \textrm{e}^{\pm i \gamma}$ 
 & $\frac{i}{15}[\cos\alpha\cos\beta \pm i\sin\alpha]
 \textrm{e}^{\pm i \gamma}$ & $0$ \\
\end{tabular}
\end{ruledtabular}
\end{table*}

\subsubsection{Spin-flipping channel}
According to Eq.~(\ref{eq.process1}), the spin-flip process is given by 
\begin{equation}
  |F\rangle \propto 
\sum_{\sigma}P_{\sigma}^{(1)}
  |\sigma\rangle\sum_{\eta}|\phi_{\eta}\rangle R(\epsilon'_{\eta})
  \langle\phi_{\eta}|\psi_{0}^{-\sigma}\rangle 
\label{eq.spin-flip}
\end{equation}
We expand $|F\rangle$ by $S_{0}^{-}|g\rangle$ and $S_{0}^{+}|g\rangle$ 
with neglecting excitations outside the core-hole site
($S_0^{\pm}\equiv S_0^{x'}\pm i S_0^{y'}$). Note that they
are orthogonal to each other and to $|g\rangle$, but not normalized,
that is, $\langle g|S_0^{-}S_0^{+}|g\rangle=
\langle\psi_0^{\downarrow}|\psi_0^{\downarrow}\rangle$ and
$\langle g|S_0^{+}S_0^{-}|g\rangle=
\langle\psi_0^{\uparrow}|\psi_0^{\uparrow}\rangle$. 
Therefore, introducing the quantity 
\begin{equation}
   f_{\sigma}^{(1)}(\omega_i) =
    \frac{1}{\langle\psi_{0}^{\sigma}|\psi_{0}^{\sigma}\rangle}
     \langle\psi_0^{\sigma}| \sum_{\eta}
     |\phi_{\eta}\rangle R(\epsilon_{\eta}')
     \langle\phi_{\eta}|\psi_0^{\sigma}\rangle, 
\label{eq.B}
\end{equation}
we have 
\begin{equation}
 |F\rangle \sim P_{\downarrow}^{(1)}f_{\uparrow}^{(1)}(\omega_i)S_0^{-}|g\rangle
             +  P_{\uparrow}^{(1)}f_{\downarrow}^{(1)}(\omega_i)S_0^{+}|g\rangle .
\label{eq.FP}
\end{equation}
Since the process for $\sigma=\uparrow$ is generally different from that for 
$\sigma=\downarrow$ in the antiferromagnetic state,
$f_{\sigma}^{(1)}(\omega_i)$ may be written as
\begin{equation}
   f_{\sigma}^{(1)}(\omega_i) = f_{0}^{(1)}(\omega_i)
   \pm \frac{1}{2}\Delta(\omega_i),
\label{eq.BD}
\end{equation}
where plus and minus signs in the second term correspond to
$\sigma=\uparrow$ and $\downarrow$, respectively.
The $\Delta(\omega_i)$, which is expressed as 
$f_{\uparrow}^{(1)}(\omega_i)-
f_{\downarrow}^{(1)}(\omega_i)$, 
is proportional to the sublattice magnetization
when it is small, since it vanishes without the antiferromagnetic
long-range order. Inserting Eq.~(\ref{eq.BD}) into Eq.~(\ref{eq.FP}),
we obtain the final expression.
For example, we have for $\alpha_f$ along the $x$ axis and $\alpha_i$ 
along the $z$ axis, 
\begin{eqnarray}
 |F\rangle &\propto& \left(\frac{2}{15}\right)if^{(1)}(\omega_i)
                     \left[M_{12} S_0^{x'} + M_{22} S_0^{y'}\right] 
\nonumber \\
           &+& \frac{1}{15}\Delta(\omega_i)
                     \left[M_{12}S_0^{y'} - M_{22} S_0^{x'}\right], 
\end{eqnarray}
where $M_{ij}$ stands for the $(i,j)$ component of the conventional
rotation matrix with the Euler angles $(\alpha,\beta,\gamma)$.\cite{Com2}
A full consideration over the polarizations leads to 
\begin{eqnarray}
  |F\rangle &\propto& 
  \left(-\frac{2}{15}\right) if_{0}^{(1)}(\omega_i)
  (\mbox{\boldmath{$\alpha$}}_{f}\times 
  \mbox{\boldmath{$\alpha$}}_{i}) \cdot {\bf S}_{0\perp}|g\rangle \nonumber\\
  &-&\left(-\frac{1}{15}\right) \Delta(\omega_i)
  (\mbox{\boldmath{$\alpha$}}_{f}\times 
  \mbox{\boldmath{$\alpha$}}_{i}) \cdot ({\bf e}_m\times {\bf S}_0)|g\rangle, \label{eq.pol-chg}
\end{eqnarray}
where ${\bf S}_{0\perp}$ represents the component perpendicular to
the direction of the staggered magnetic moment, and $\rm{\bf e}_m$
represents the unit vector along the direction of the sublattice
magnetization.

\subsubsection{Spin-conserving channel}
According to Eq.~(\ref{eq.process1}), the spin-conserving process is given by 
\begin{equation}
 |F\rangle \propto 
\sum_{\sigma} P_{\sigma}^{(0)}
  |\sigma\rangle\sum_{\eta}|\phi_{\eta}\rangle R(\epsilon'_{\eta})
  \langle\phi_{\eta}|\psi_{0}^{\sigma}\rangle.
\end{equation}
We expand $|F\rangle$ by $|g\rangle$ and $S_0^{z'}|g\rangle$ by neglecting
the excitations outside the core-hole site.
Note that $S_{0}^{z'}|g\rangle$ is not orthogonal to 
$|g\rangle$ nor normalized. 
Let $|\psi_1\rangle$ and $|\psi_2\rangle$
be $|g\rangle$ and $S_0^{z'}|g\rangle$, respectively. 
Then the overlap matrix $[\hat{\rho}]_{i,j}\equiv 
\langle \psi_i|\psi_j\rangle$ is given by 
\begin{equation}
 \hat{\rho} = \left( \begin{array}{cc}
                       1 & \langle S_0^{z'}\rangle \\
  \langle S_0^{z'}\rangle & \frac{1}{4}
                     \end{array} \right) .
\end{equation}
We project onto these states by operating 
$\sum_{i,j}|\psi_i\rangle(\hat{\rho}^{-1})_{i,j}\langle\psi_j|$.
For the channel preserving the direction of the polarization 
during the scattering process, we have
\begin{equation}
 |F\rangle \propto \left(\frac{2}{15}\right)
  (\mbox{\boldmath{$\alpha$}}_{f}\cdot 
  \mbox{\boldmath{$\alpha$}}_{i})\left[ 
  f_0(\omega_i)|g\rangle + \Delta(\omega_i){\bf e}_m \cdot{\bf S}_0|g\rangle
  \right].
\label{eq.pol-conserv}
\end{equation}
Similarly, for the scattering
channel changing the direction of the polarization during the process, 
by using $P_{\uparrow}^{(0)}=-P_{\downarrow}^{(0)}$, 
we obtain
\begin{eqnarray}
  |F\rangle &\propto& 
   \left(-\frac{1}{15}\right) i\frac{1}{2}\Delta(\omega_i)
  (\mbox{\boldmath{$\alpha$}}_{f}\times 
  \mbox{\boldmath{$\alpha$}}_{i}) \cdot {\bf e}_m|g\rangle \nonumber\\
  &+&\left(-\frac{2}{15}\right) if_{0}^{(1)}(\omega_i)
  (\mbox{\boldmath{$\alpha$}}_{f}\times 
  \mbox{\boldmath{$\alpha$}}_{i}) \cdot {\bf S}_{0\parallel}|g\rangle,
\label{eq.spin-conserv}
\end{eqnarray}
where ${\bf S}_{0\parallel}$ represents the component parallel to
the direction of the staggered magnetic moment.

\subsubsection{Elastic scattering}
The amplitude of elastic scattering is given by $\langle g|F\rangle$. 
The first term of Eq.~(\ref{eq.pol-conserv})
gives a contribution independent of the magnetic order,
while the second term of Eq.~(\ref{eq.pol-conserv})
gives a contribution proportional to $|{\bf m}|^2$,
since $\Delta(\omega_i)$ is proportional to $|{\bf m}|$.
Here ${\bf m}$ stands for the sublattice magnetization.
Both terms in Eq.~(\ref{eq.spin-conserv}) give the contributions 
proportional to $(\mbox{\boldmath{$\alpha$}}_{f}\times 
\mbox{\boldmath{$\alpha$}}_{i}) \cdot {\bf m}$,
which is consistent with the formula given by Hannon \textit{et. al}.
\cite{Hannon1988}

\subsubsection{Remarks}
Here, it is interesting to compare our result derived on the basis
of the projection method with other well-known results;
one is the far-off-resonance condition that 
$|\omega_i-\epsilon_{\rm core}|\gg |\epsilon'_{\eta}-\epsilon_g|$,
and another is the large limit of $\Gamma$, 
which is called as the fast collision approximation (FCA).
\cite{Ament2007,Ament2009,Haverkort2010}
In both latter conditions, we could factor out $R(\epsilon'_{\eta})$ 
from the summation over $\eta$ in Eq.~(\ref{eq.B}). Then,  
using the closure relation of $|\phi_{\eta}\rangle$,
we immediately obtain $\Delta(\omega_i)=0$.
The presence of $\Delta(\omega_i)$ is a hallmark of a second-order 
process that the x ray could recognize the long-range order in the 
scattering process, 
contrast with neutron scattering.
In the present case, however, $\Delta(\omega_i)$  is estimated to be 
quite small, since the life-time broadening width is rather large
at the Ir L-edge.\cite{Krause1979}
By neglecting $\Delta(\omega_i)$, 
Eqs. (\ref{eq.pol-chg}) and (\ref{eq.spin-conserv}) 
are summarized into an expression, which is 
similar to that for the undoped cuprates, 
\cite{Ament2007,Ament2009,Haverkort2010,Igarashi2012,Com3}
as
\begin{equation}
  |F\rangle \propto
  \left(-\frac{2}{15}\right) if_{0}^{(1)}(\omega_i)
  (\mbox{\boldmath{$\alpha$}}_{f}\times 
  \mbox{\boldmath{$\alpha$}}_{i}) \cdot {\bf S}_0|g\rangle.
\label{eq.scatop}
\end{equation}
Note that when both $f_{0}^{(1)}(\omega_i)$ and $\Delta(\omega_i)$
are numerically relevant, their $\omega_i$ dependence might be a 
intriguing feature.
However, once $\Delta(\omega_i)$ is neglected as in the present case, 
we do not have to evaluate
the value of $f_{0}^{(1)}(\omega_i)$, since RIXS cannot tell about
the absolute magnitude of the intensity.

\section{\label{sect.4}Analysis of RIXS spectra 
from S\lowercase{r}$_2$I\lowercase{r}O$_4$}
\begin{figure}
\includegraphics[width=8.0cm]{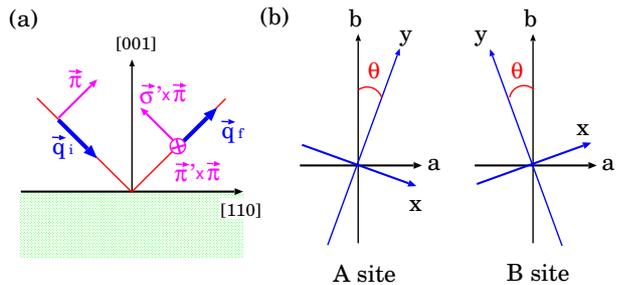}%
\caption{\label{fig.geometry} (Color online)
(a) Geometry of 90$^{\circ}$ scattering. The scattering plane is
perpendicular to the $ab$ plane and intersects the $ab$ plane with
the $[110]$ direction.
(b) Local coordinate frames of the two sublattices, 
which are rotated by angle $\pm \theta$ around the $c$ axis.
}
\end{figure}

We consider the specific case of a 90$^{\circ}$ scattering angle.
The scattering plane is perpendicular to the IrO$_2$ plane and intersects
the $ab$ plane with the $[110]$ direction, as illustrated
in Fig.~\ref{fig.geometry}(a). The incident x ray is 
assumed to have the $\pi$ 
polarization. 
Since $\omega\sim 11.2$ keV and $|{\bf q}_i|\sim 5.7$ $\textrm{\AA}^{-1}$
at the Ir $L_3$ edge, only a few degrees of tilt of the scattering plane
could sweep the entire Brillouin zone. 

The scattering operator $Z^{(1)}({\bf q})$ is given by summing up the 
amplitude with multiplying $\exp(i{\bf q}\cdot{\bf r}_{j})$ 
at each Ir site ${\bf r}_{j}$,
\begin{equation}
 Z^{(1)}({\bf q}) \equiv \frac{1}{\sqrt{N}}\sum_{j} 
 (\mbox{\boldmath{$\alpha$}}_{f}\times 
\mbox{\boldmath{$\alpha$}}_{i})\cdot {\bf S}_{j}{\rm e}^{-i \textbf{q}
\cdot \textbf{r}_{j}},
\end{equation}
where ${\bf q}\equiv{\bf q}_i-{\bf q}_f$ is the momentum transfer.
Note that the local coordinate frames defining spin operators are different 
between the A and B sites, as illustrated in Fig.~\ref{fig.geometry}(b).
Evaluating $\mbox{\boldmath{$\alpha$}}_{f}\times 
\mbox{\boldmath{$\alpha$}}_{i}$ in the local coordinate frame,
we have $Z^{(1}({\bf q})$ for ${\bf q}$ inside the first MBZ, 
\begin{eqnarray}
 Z^{(1)}({\bf q})
& =& -\frac{1}{2}\left[\cos\theta\, Q_{+}^{x}({\bf q})
                    -\sin\theta\, Q_{-}^{x}({\bf q}) \right. \nonumber \\
               &&  \left.   +\cos\theta\, Q_{+}^{y}({\bf q})
                    +\sin\theta\, Q_{-}^{y}({\bf q})\right]
   +\frac{1}{\sqrt{2}}Q_{+}^{z}({\bf q}),
\label{eq.Zs}
\end{eqnarray}
in the $\sigma'\times\pi$ channel, and 
\begin{eqnarray}
 Z^{(1)}({\bf q})
 &=& -\frac{1}{\sqrt{2}}\left[\cos\theta\, Q_{+}^{x}({\bf q})
                    +\sin\theta\, Q_{-}^{x}({\bf q}) \right. \nonumber \\
   &&             \left.    -\cos\theta\, Q_{+}^{y}({\bf q})
                  +\sin\theta\, Q_{-}^{y}({\bf q})\right].
\label{eq.Zp}
\end{eqnarray} 
in the $\pi'\times\pi$ channel.
The scattering operators for ${\bf q}$ outside the first MBZ
are given by replacing $Q_{\pm}^{\mu}({\bf q})$ with
$Q_{\mp}^{\mu}([{\bf q}])$.

The RIXS intensity is proportional to the correlation functions for
these scattering operators, 
\begin{equation}
 I 
\equiv 
%\propto 
W(q_f\alpha_f;q_i\alpha_i) \propto \int_{-\infty}^{\infty}
 \langle Z^{(1)}({\bf q},t)Z^{(1)}(-{\bf q},0)
 \rangle {\rm e}^{i\omega t}{\rm d}t.
\label{eq.corZ}
\end{equation}
The insertion of Eqs.~(\ref{eq.Zs}) and (\ref{eq.Zp}) into Eq.~(\ref{eq.corZ})
leads to the expression for ${\bf q}$ inside the first MBZ 
\begin{equation}
 I 
%= 
\propto 
\left\{ \begin{array}{ll}
     \frac{\cos^2\theta\, R_{+}^{yy}({\bf q},\omega)
   + \sin^2\theta\, R_{-}^{yy}({\bf q},\omega)
   + 2 R_{+}^{zz}({\bf q},\omega)}{4},
 & {\rm for}\ \sigma'\times\pi, \\
     \frac{\cos^2\theta\, R_{+}^{yy}({\bf q},\omega)
   +\sin^2\theta\, R_{-}^{yy}({\bf q},\omega)}{2},
 & {\rm for}\ \pi'\times\pi, \end{array} \right. .
\label{eq.W}
\end{equation} 
We have neglected $R_{\pm}^{xx}({\bf q},\omega)$, since it is a higher order 
of $1/S$. To extend the expression to outside the first MBZ,
$R_{+}^{yy}({\bf q},\omega)$, $R_{+}^{zz}({\bf q},\omega)$ and 
$R_{-}^{yy}({\bf q},\omega)$ are replaced by 
$R_{-}^{yy}([{\bf q}],\omega)$ and $R_{-}^{zz}([{\bf q}],\omega)$ and
$R_{+}^{yy}([{\bf q}],\omega)$, respectively.
Note that the $\sin^2\theta$-terms give the \emph{antibonding}
contribution for ${\bf q}$ \emph{inside} the first MBZ, 
which diverges at $\omega=0$ with ${\bf q}\to (0,0)$. 
This unusual contribution may be interpreted as a reflection of the weak 
ferromagnetism.

\begin{figure}
\includegraphics[width=8.0cm]{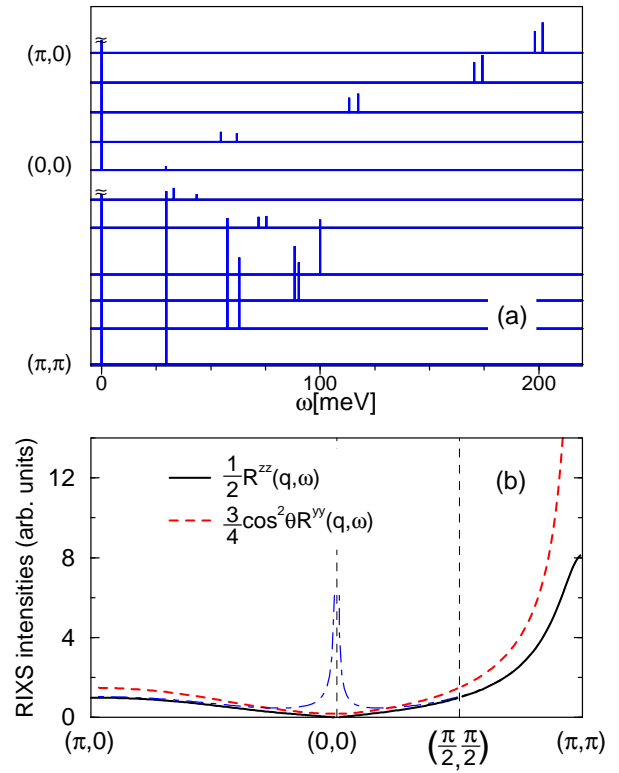}%
\caption{\label{fig.rixs} (Color online)
(a) RIXS spectra as a function of $\omega$ for ${\bf q}$ 
along symmetry lines evaluated for $\theta=\pm 11^{\circ}$.
Vertical bars represent the $\delta$-function peaks with heights 
proportional to their intensities. The height for the peaks at $\omega=0$ 
is divergent, and are cut to be finite on the figure.
(b) Peak intensities of $\frac{1}{2}R^{zz}({\bf q},\omega)$ 
(black solid line)
and $\frac{3}{4}\cos^2\theta R^{yy}({\bf q},\omega)$ (red broken line)
for $\theta=0$. The (blue) broken-dotted line shows the intensity 
added by the $\sin^2\theta$ terms for $\theta=\pm 11^{\circ}$, 
which makes the curve 
deviate from the curve at $\theta=0$ for ${\bf q}$ only close to $(0,0)$.
}
\end{figure}

Figure \ref{fig.rixs} shows the numerical results with
the same parameter values as for the correlation function.
Panel (a) shows the RIXS spectra as a function of 
$\omega$ for ${\bf q}$ along the symmetry lines, and panel (b) shows
the intensities of two peaks.
The intensities from the $\sigma'$ and $\pi'$ polarization channels are 
summed up.
At ${\bf q}=(0,0)$, the intensity of the peak diverges at $\omega=0$
due to the weak ferromagnetism ($\sin^2\theta$-term), while that of another 
peak is quite small at $\omega=29$ meV. 
The effect of the weak ferromagnetism is limited very close to the $\Gamma$
point. 
At ${\bf q}=(\pi,\pi)$, the intensity of the peak also diverges 
at $\omega=0$ due to the antiferromagnetic order, while that of another peak
is rather large at $\omega=29$ meV. 

\section{\label{sect.5}Concluding remarks}
We have studied the magnetic excitations in Sr$_2$IrO$_4$ on the basis of
the Heisenberg model with isotropic exchange couplings and small anisotropic 
terms. Solving the coupled equations of motion for the Green's functions
within the spin-wave approximation, we have found that two modes emerge 
with slightly different energies. Introducing the bonding and antibonding 
combinations of spin operators at A and B sites, we have considered the 
correlation functions for these operators. We have found that 
the correlation functions with the $y$ and $z$ spin-components 
are composed of a single $\delta$-function peak with different energies 
corresponding to each mode. We have analyzed the second-order RIXS process with the 
assumption that the excitations are confined on the core-hole site,
and have obtained the expression for the local scattering operator composed of
the term consistent with the FCA as well as a term existing only in the 
broken symmetric phase.
The latter is, however, expected to be quite small in Sr$_2$IrO$_4$, 
since the life-time broadening width at the $L$ edge of Ir is rather large.
Using the scattering operator, the RIXS intensity has been expressed by 
a sum of the correlation functions with two spin components.
Having evaluated the formula, we have demonstrated that the spectra are
composed of two peaks originated from the split modes. 
Such two-peak structures have not been observed in the RIXS experiments. 
\cite{J.Kim2012,Ament2011}
We hope that the present analysis may help to verify the mode splitting
in the experiments with improving the instrumental energy resolution.
\cite{MorettiSala2013}

Here, we comment on the effect of $\Delta(\omega_i)$ on the RIXS spectrum, which becomes relevant when the core-hole lifetime broadening $\Gamma$ is small. 
It then requires a reliable evaluation of the coefficients $f_{0}^{(1)}(\omega_i)$ and $\Delta(\omega_i)$ to calculate the RIXS intensity.
In our previous work, we have confirmed that analysis utilizing a small cluster works well in evaluating the coefficients with moderate accuracy for cuprates, which have revealed that the RIXS intensity showed a characteristic q-dependence for small $\Gamma$.\cite{Igarashi2012} 
However, such evaluation for the present case is very difficult because the magnitude of the exchange coupling between the third neighbors remains significant in Sr$_2$IrO$_4$, which requires an  analysis for a larger cluster. 
An analysis with high accuracy in this direction will be an intriguing future work.

The present study is based on the localized electron picture,
which works well on the magnetic excitations in the strong coupling limit. 
\cite{Anderson1959}
However, other peak structures have been observed around the region of 
$0.4 \sim 0.6$ eV in the RIXS experiment, which could be attributed to 
the excitations from $j_{\textrm{eff}}=1/2$ to $3/2$ 
multiplets.\cite{J.Kim2012,Ament2011}
Since the Mott-Hubbard gap is estimated as $\sim 0.4$ eV from the
optical absorption spectra,\cite{Kim2008,Moon2009} this energy region
also coincides with the energy 
continuum of the electron-hole pair creation.
In such a situation, it may make sense to consider the spectra
from the itinerant electron picture in order to obtain a coherent picture
of RIXS spectra. Such study based on the Hartree-Fock and RPA approximations
is under progress.\cite{Igarashi2013-1}

\begin{acknowledgments}
We are grateful to M. Yokoyama and K. Ishii for fruitful discussions.
This work was partially supported by a Grant-in-Aid for Scientific Research
from the Ministry of Education, Culture, Sports, Science and Technology
of the Japanese Government.
\end{acknowledgments}

\appendix 
\section{Symmetry relations among the Green's functions}
We consider the Green's function defined by
\begin{equation}
 G_{AB}(\omega)=-i\int\langle T[A(t)B(0)]\rangle 
     {\rm e}^{i\omega t}{\rm d}t,
\end{equation}
where $A$ and $B$ are boson operators. It is expressed in the spectral
representation as
\begin{equation}
 G_{AB}(\omega) = \sum_{n} \left\{
  \frac{\langle g|A|n\rangle\langle n|B|g\rangle}{\omega-E_n+E_g+i\delta}
 -\frac{\langle g|B|n\rangle\langle n|A|g\rangle}{\omega+E_n-E_g-i\delta}
  \right\},
\end{equation}
where $|n\rangle$ stands for the eigenstate of the Hamiltonian with energy 
$E_n$, and $|g\rangle$ the ground state with energy $E_g$.
It is easily proved from this expression that
\begin{equation}
 G_{B^{\dagger}A^{\dagger}}(\omega)=G_{AB}(\omega),\quad
 G_{A^{\dagger}B^{\dagger}}(\omega)=G_{AB}(-\omega).
\end{equation}
Hence we obtain the relations between the Green's functions of
Holstein-Primakoff bosons by replacing A by one of $a({\bf k})$, 
$a^{\dagger}({-\bf k})$, $b({\bf k})$, $b^{\dagger}({-\bf k})$,
and B by one of $a(-{\bf k})$, $a^{\dagger}({\bf k})$, $b(-{\bf k})$, 
$b^{\dagger}({\bf k})$.
In addition, since the Hamiltonian is invariant with exchanging $a$ and $b$
as well as $a^{\dagger}$ and $b^{\dagger}$,
the Green's functions remain the same forms by such exchange.

\bibliography{paper}

\begin{thebibliography}{37}
\bibitem[{Com({\natexlab{a}})}]{Com1}
\bibinfo{note}{In the $j_{\rm eff}=\frac{1}{2}$ manifold,
  $h_{\uparrow}^{\dagger}$ and $h_{\downarrow}^{\dagger}$ are defined by
  $h_{\uparrow}^{\dagger}|{\rm vac}\rangle=d_{\downarrow}|0\rangle$,
  $h_{\downarrow}^{\dagger}|{\rm vac}\rangle=-d_{\uparrow}|0\rangle$, with
  $|{\rm vac}\rangle=d_{\uparrow}^{\dagger}d_{\downarrow}^{\dagger}|0\rangle$,
  where $d_{\sigma}$ is the annihilation operator of electron with the Kramers'
  doublet specified by $\sigma$.}
\bibitem[{Com({\natexlab{c}})}]{Com3}
\bibinfo{note}{For undoped cuprates, $\alpha_i$ and $\alpha_f$ in Eq.
  (\ref{eq.scatop}) are replaced by those projected onto the $ab$ plane.}
\end{thebibliography}

\end{document}